\documentclass[final,3p,times,twocolumn,authoryear]{elsarticle}
\usepackage{amssymb}
\usepackage{color}
\usepackage{bm}
\usepackage{url}

\begin{document}

\begin{frontmatter}



\title{Density functional theory}


\author[inst1]{Yusuke Nomura}

\affiliation[inst1]{organization={Department of Applied Physics and Physico-Informatics, Keio University},%Department and Organization
            addressline={3-14-1 Hiyoshi, Kohoku-ku}, 
            city={Yokohama},
            postcode={223-8522}, 
            country={Japan}}

\author[inst2]{Ryosuke Akashi}

\affiliation[inst2]{organization={Department of Physics, University of Tokyo},%Department and Organization
            addressline={Hongo 7-3-1,}, 
            city={Bunkyo},
            postcode={113-8656}, 
            country={Japan}}

\begin{abstract}
Density functional theory (DFT) is an essential building block for modern theoretical physics, chemistry, and engineering, especially those concerning electronic properties. Through decades of development, various program packages for first-principles electronic structure calculation are now available. Their sophisticated interfaces allow users to apply DFT to actual systems, even without knowing the theory. It is hence becoming more and more important to recall the fundamentals of how DFT enables accurate calculations. 
This article attempts to provide such knowledge with a minimal overview of DFT---its basic foundation, relations to observable electronic and nuclear dynamical properties, and some of its cutting-edge applications.
\end{abstract}

\begin{keyword}
Density functional theory \sep Electronic structure \sep Nuclear dynamics
\end{keyword}

\end{frontmatter}


\section*{Key points/Objectives}
%\begin{breakitembox}[l]{Key points}
\begin{itemize}
    \item Theoretical foundation of density functional theory is reviewed.
    \item Relations to experimentally observable quantities are highlighted.
    \item Modern applications of density functional theory are introduced.
\end{itemize}

%\end{breakitembox}

\section{Introduction} \label{sec_overview}

The motion of the electrons in atoms, molecules, and solids is described by the Schr$\ddot{\rm o}$dinger (or Dirac) equation. 
The computational complexity to obtain the exact solution grows exponentially with the number of electrons $N$. 
Calculating exact ground-state wave function is prohibitively expensive when $N$ is more than a few dozen.  
Therefore, it has been a great challenge to develop a powerful and accurate numerical method to calculate electronic structures.

There are mainly two numerical approaches. 
One is the wave function theory. 
It directly deals with the many-body wave function itself and attempts to find good approximations to the exact wave function. 
The other is the density functional theory (DFT), for which we will give a brief review.  
By employing the electron density $\rho({\bf r})$ (a function of three coordinate variables) as the fundamental variable instead of the many-body wave function (a function of $3N$ coordinate variables), DFT has drastically reduced the computational cost.
Therefore, DFT is the most widely used method for electronic structure calculations of solids.

Here, we briefly review the fundamental and practical aspects of DFT. 
Sec.~\ref{sec:formalism} reviews the Hohenberg-Kohn theorem and the Kohn-Sham equation, which gives a foundation of DFT. 
We also describe several extensions of DFT. 
In DFT, as described below, the exchange-correlation functional is one of the most fundamental quantities. 
Sec.~\ref{sec:XC} is devoted to a discussion of approximations to the exchange-correlation functional. 
Sec.~\ref{sec:electronic_structure} presents several topics related to electronic structure calculations.
In Sec.~\ref{sec:lattice}, we show that DFT can also be used to calculate the properties of atomic vibrations.
Sec.~\ref{sec:practical} provides a practical guide of DFT. 
Finally, we give a summary in Sec.~\ref{sec:summary}.

\section{Basic formalism}
\label{sec:formalism}

We start from the full non-relativistic Hamiltonian, in which electrons and atomic nuclei interact with each other. 
Because the masses of nuclei are much heavier than that of electrons, the kinetic energies of nuclei are usually much smaller than that of electrons. 
Then, to a good approximation, we can treat electron and nuclear dynamics separately.
DFT is a theory for dealing with the electronic part under this approximation.
In this section, we first briefly discuss the Born-Oppenheimer approximation~\citep{Born_Oppenheimer}, which is the most widely used framework to derive separate equations for the electron and nuclear dynamics. 
Then, we discuss the Hohenberg-Kohn theorem~\citep{Hohenberg_Kohn} and the Kohn-Sham equation~\citep{Kohn_Sham}, which form the basis of DFT.

\subsection{Born-Oppenheimer approximation}
The Hamiltonian for interacting electrons and nuclei reads 
\begin{eqnarray}\label{Eq:H_all}
\hat{\mathcal H} = -\sum_{I}\frac{\hbar^{2}}{2M_{I}}\frac{\partial^{2}}{\partial{\bf R}_{I}^{2}}
-\sum_{i}\frac{\hbar^{2}}{2m} \frac{\partial^{2}}{\partial{\bf r}_{i}^{2}} 
+V\bigl(\{ {\bf r} \}, \{ {\bf R}\}  \bigr),
\end{eqnarray}
where 
\begin{eqnarray}
\hspace{-0.6cm}
V(\{ {\bf r} \} , \{ {\bf R} \} ) =  \sum_{i <  j} \frac{e^2}{|{\bf r}_{i} \! - \! {\bf r}_{j}|}
 -  \sum_{i,I} \frac{Z_{I}e^{2}}{|{\bf r}_{i} \! - \! {\bf R}_{I}|}
 +  \sum_{I <  J} \frac{Z_{I}Z_{J} e^2}{|{\bf R}_{I} \! - \! {\bf R}_{J}|}  \nonumber 
\end{eqnarray}
with $\{ {\bf r} \}$ and $\{ {\bf R}  \}$ being the set of electron and nucleus coordinates, respectively,
and $i$ and $I$ labeling electrons and nuclei. 
The Born-Oppenheimer approximation makes use of the large difference in the mass between the nuclei and electrons. 
Within this approximation, the electron and nuclear motions are treated separately, and the wave function of electrons and nuclei is given by a product of the electron part $\Psi_{\rm e}$ and the nuclear part $\Phi_{\rm n}$.
In the following, we briefly describe the equations obtained from the Born-Oppenheimer approximation.

If we neglect the nuclear kinetic energy, which is much smaller than that of electrons, from the Hamiltonian, we obtain the electron Schr$\ddot{\rm o}$dinger equation: 
\begin{eqnarray}
\label{Eq:BO_electron}
\biggl [ - \sum_{i}\frac{\hbar^{2}}{2m} \frac{\partial^{2}}{\partial{\bf r}_{i}^{2}} 
 +  V \bigl(\{ {\bf r} \}, \{ {\bf R} \} \bigr) \biggr] \Psi_{\rm e} \bigl( \{ {\bf r} \}  \ \!   | \  \!   \{ {\bf R} \} \bigr ) \nonumber \\
 =   E \bigl( \{ {\bf R} \} \bigr ) \Psi_{\rm e} \bigl (\{ {\bf r} \}  \ \!   | \  \!   \{ {\bf R} \} \bigr ).
\end{eqnarray}
Here, the equation is solved for fixed nuclear positions. 
The nuclear dynamics with a smaller energy scale are treated separately from the electron dynamics. 
The Schr$\ddot{\rm o}$dinger equation for the nuclear part is given by
\begin{eqnarray}
\label{Eq:lattice}
\biggl[-\sum_{i}\frac{\hbar^{2}}{2M_{I}} \frac{\partial^{2}}{\partial{\bf R}_{I}^{2}} 
+E\bigl( \{ {\bf R} \} \bigr ) \biggr] \Phi_{\rm n} \bigl( \{ {\bf R} \} \bigr)  = \varepsilon \Phi_{\rm n} \bigl (\{ {\bf R} \} \bigr), 
\end{eqnarray}
where the total energy of the electron part $E\bigl( \{ {\bf R} \} \bigr )$ serves as a potential for the nuclear dynamics (Born-Oppenheimer energy surface). 
$\varepsilon$ is the total energy of the electron-nucleus coupled system.
In the following, we explain the idea of DFT for solving the electronic part [Eq.~(\ref{Eq:BO_electron})]. 
We also discuss the nuclear dynamics in Sec.~\ref{sec:lattice}.

%Then, the wave function can be expressed by a product of the electron part $\Psi_{\rm e}$ and the lattice part $\Phi_{\rm n}$ as 
%\begin{eqnarray}
%\Phi \bigl(  \{ {\bf r} \} ,\{ {\bf R} \},t \bigr ) \simeq \Phi_{\rm n}\bigl (\{ {\bf R} \} \bigr )\Psi_{\rm e} \bigl ( \{ {\bf r} \}  \ \!   | \  \!   \{ {\bf R} \} \bigr )e^{-i\varepsilon t/\hbar}.
%\end{eqnarray}
%Substituting this form in Eq.~(\ref{Eq:td_schr}), we obtain separate equations for electron and lattice dynamics:  

\subsection{Hohenberg-Kohn theorem and Kohn-Sham equation}
\label{sec_HK_KS}

The electronic part of the Hamiltonian is given by
\begin{eqnarray}
\hspace{-0.5cm}
\hat{\mathcal H}_{\rm e} = 
\sum_{i} \Biggl( - \frac{\hbar^{2}}{2m} \frac{\partial^{2}}{\partial{\bf r}_{i}^{2}} 
-  \sum_{I} \frac{Z_{I}e^{2}}{|{\bf r}_{i}-{\bf R}_{I}|}\Biggr) + \sum_{i<j}  \frac{e^2}{|{\bf r}_{i}-{\bf r}_{j}|}.
\end{eqnarray}
As is mentioned in Sec.~\ref{sec_overview}, when the number of electrons is more than a few dozen, it is prohibitively demanding to obtain the exact ground-state wave function. 

In DFT, the electron density $\rho({\bf r})$ is used as the fundamental variable instead of the many-body wave function.
The electron density is much more tractable than the many-body wave function because it depends on only three coordinate variables, regardless of the number of electrons.
This approach has been justified by the Hohenberg-Kohn theorem.

The theorem first states one-to-one correspondence between the ground-state electron density and the external potential. 
Therefore, once the ground-state electron density is known, the external potential is determined uniquely.  
%Then, in principle, all physical properties can be calculated from the electron density. 
Then, in principle, physical properties associated with the ground-state wave function are unambiguously determined by the electron density. 
In particular, the kinetic and electron-electron interaction energies of the ground state can be expressed as universal functionals of the electron density ($E_{\rm kin} [ \rho]$ and $E_{\rm ee} [\rho]$, respectively). 
The term ``universal'' indicates that the forms of the functionals do not depend explicitly on the external potential.

The theorem also gives a variational principle. 
When we define the energy functional 
\begin{eqnarray}
E_\varv[ \rho]  =   E_{\rm kin} [ \rho]  +  \int \!   \rho({\bf r}) \varv({\bf r}) d{\bf r}   + E_{\rm ee} [\rho] 
\end{eqnarray}
for some external potential $\varv({\bf r})$, the functional satisfies the inequality 
\begin{eqnarray}
\label{eq_variational}
E_\varv[ \rho]  \geq  E_\varv[\rho_0] =  E_0 , 
\end{eqnarray}
where $\rho_0$ ($E_0$) is the ground-state electron density (energy) under the potential $\varv({\bf r})$, respectively. 
Therefore, the ground-state electron density can be obtained by searching for the electron density that minimizes the energy functional $E_\varv[ \rho]$.

Although the variational principle in Eq.~(\ref{eq_variational}) looks simple, the major problem is that the exact forms of the functionals, $E_{\rm kin} [ \rho]$ and $E_{\rm ee} [\rho]$, are unknown. 
For this problem, Kohn and Sham have proposed an idea of introducing ``orbitals'' to approximate the kinetic energy functional $E_{\rm kin} [ \rho]$.
This has opened up a way to perform DFT calculations with sufficient accuracy for practical use.
Therefore, most of the modern DFT implementations employ the Kohn-Sham scheme; thus, we discuss this scheme in more detail below.  
As for the direct variational approach (orbital-free DFT), which is less accurate than the Kohn-Sham approach but has the advantage of being much faster, the current research focus is to construct accurate kinetic energy functionals.
One strategy is to reproduce the Kohn-Sham kinetic energy as accurately as possible.

The Kohn-Sham scheme introduces an auxiliary non-interacting system that is designed to give the same electron density as that of the interacting system.  
The non-interacting system is described by single-particle Schr$\ddot{\rm o}$dinger equation (so called Kohn-Sham equation): 
\begin{eqnarray}
\label{eq:KS2}
\left[  - \frac{\hbar^{2}}{2m}  \frac{\partial^{2}}{\partial{\bf r}^{2}} + \varv_{\rm eff} ({\bf r }) \right]   \phi_i ({\bf r}) = \varepsilon_{i} \phi_i ({\bf r}),
\end{eqnarray}
where $\varv_{\rm eff} ({\bf r })$ is an effective potential which is determined by Eq.~(\ref{eq:veff}), $\phi_i$ is the Kohn-Sham state, and $\varepsilon_{i}$ is the Kohn-Sham energy eigenvalue. 
The electron density is given by 
\begin{eqnarray}
\label{eq:density}
\rho({\bf r}) =  \sum_{i=1}^{\rm occ.} | \phi_i({\bf r}) | ^2,
\end{eqnarray}
where the summation runs over occupied Khon-Sham states.

In the Kohn-Sham scheme, to reproduce the true electron density with Eq.~(\ref{eq:density}), the effective potential $\varv_{\rm eff}({\bf r})$ is determined as follows: 
We first introduce the kinetic energy functional 
\begin{eqnarray}
E_{\rm kin}^{\rm KS}[\rho] =  \sum_{i=1}^{\rm occ.}  \int    
\phi^*_i ({\bf r})  \left( - \frac{\hbar^{2}}{2m}  \frac{\partial^{2}}{\partial{\bf r}^{2}}  \right) \phi_i ({\bf r})d {\bf r}
\end{eqnarray}
and the Hartree energy functional 
\begin{eqnarray}
E_{\rm H} [\rho] =  \frac{e^2}{2}  \int \! \!  \!  \int    
  \frac{\rho ({\bf r})  \rho ({\bf r'})} {| {\bf r} - {\bf r}'| }   d {\bf r}  d{\bf r'}.
\end{eqnarray}
of the Kohn-Sham system. 
Next, we introduce so called exchange-correlation functional, whose definition is given by the sum of corrections to the exact kinetic and electron-electron interaction energies of the original interacting system: 
\begin{eqnarray}
E_{\rm xc} [\rho] = \bigl ( E_{\rm kin} [\rho] - E_{\rm kin}^{\rm KS} [\rho]  \bigr) 
+ \bigl( E_{\rm ee} [\rho] - E_{\rm H} [\rho] \bigr).   
\end{eqnarray}
Then, one can show that $\varv_{\rm eff}({\bf r})$ should be given by 
\begin{eqnarray}
\label{eq:veff}
\varv_{\rm eff} ({\bf r} ) \!\! &=& \!\! \varv({\bf r}) +  e^2 \!  \int  \!  \frac{ \rho ({\bf r'})} {| {\bf r} - {\bf r}'| }   d{\bf r'} + \varv_{\rm xc} (\bf r), \\
\varv_{\rm xc} (\bf r) \!\!  &=& \!\! \frac{\delta E_{\rm xc} [\rho] } {\delta \rho ({\bf r})}.  
\end{eqnarray}

One can see that Eqs. (\ref{eq:KS2}), (\ref{eq:density}), and (\ref{eq:veff}) form self-consistent equations.  
Therefore, in practical DFT calculations based on the Kohn-Sham scheme, the self-consistent equations are solved iteratively with some approximate form of the exchange-correlation functional (see Sec.~\ref{sec:XC} for the detail).

\subsection{Extensions}
\label{sec:extension}
The ground states are sometimes not sufficiently characterized only by the charge density distribution. This occurs when any quantities, generally represented by the 1-reduced density matrix, are involved
\begin{eqnarray}
\rho^{(1)}_{\sigma\sigma'}({\bf r}, {\bf r}')
=
\langle
\hat{\psi}^{\dagger}_{\sigma}({\bf r})
\hat{\psi}_{\sigma'}({\bf r}')
\rangle
.
\end{eqnarray}
$\langle \hat{\mathcal O} \rangle$ denotes the expectation value of the operator $\hat{\mathcal O}$.
When the system is subjected to external fields or exhibits spontaneous symmetry breaking, its nontrivial components, such as spin density ${\bf m}({\bf r})=\sum_{\sigma\sigma'}{\bm \sigma}_{\sigma \sigma'}\rho^{(1)}_{\sigma\sigma'}({\bf r}, {\bf r}) \ \ [{\bm \sigma}=(\sigma_{x}, \sigma_{y}, \sigma_{z})]$, may become nonzero. 
The framework of DFT can be extended for such situations. Namely, we can introduce $\rho^{(1)}_{\sigma\sigma'}({\bf r}, {\bf r}')$ as fundamental variables in addition to the charge density distribution $\rho({\bf r})$. A parallelism to the Hohenberg-Kohn theorem holds: The ground state is unambiguously characterized by $\rho({\bf r})$ and combination of $\rho^{(1)}_{\sigma\sigma'}({\bf r}, {\bf r}')$. The Kohn-Sham equation, which reproduces the identical $\rho$ and $\rho^{(1)}$ in the interacting system, can be constructed with complementary exchange-correlation potentials $\varv_{\rm xc}^{\sigma\sigma'}({\bf r},{\bf r}')=\delta E_{\rm xc}[\rho, \rho^{(1)}]/\delta \rho^{(1)}_{\sigma\sigma'}({\bf r}, {\bf r}')$. 
Such an extension has been first accomplished for the spin density~\citep{vonBarth-Hedin1972}. Calculations based on the spin DFT is now widely in practice for describing the electron spin distribution of materials. More generally, the 1-reduced density matrix functional theory has also been developed, which treats all the $\rho^{(1)}_{\sigma\sigma'}({\bf r}, {\bf r}')$ components as additional variables.

%The 1-reduced density matrix functional theory thus treats all the $\rho^{(1)}_{\sigma\sigma'}({\bf r}, {\bf r}')$ components as additional variables. As its special case, the spin density functional theory, with ${\bf m}({\bf r})=\sum_{\sigma\sigma'}{\bm \sigma}_{\sigma \sigma'}\rho^{(1)}_{\sigma\sigma'}({\bf r}, {\bf r}) \ \ [{\bm \sigma}=(\sigma_{x}, \sigma_{y}, \sigma_{z})]$ being the variables, has now been widely in practical use. 

The DFT framework is further extended for degrees of freedom other than the electronic charge and spin. Here we list some representative examples. In the current density functional theory, the current density ${\bm j}({\bf r})=\sum_{\sigma}\left\langle\frac{-e\hbar}{2mi}\left[\hat{\psi}^{\dagger}_{\sigma}({\bf r})\nabla\hat{\psi}_{\sigma}({\bf r})-[\nabla\hat{\psi}^{\dagger}_{\sigma}({\bf r})]\hat{\psi}_{\sigma}({\bf r})\right]\right\rangle$ is treated as a variable. 
Inclusion of the off-diagonal order variable $\chi_{\sigma\sigma'}({\bf r},{\bf r}')=\langle\hat{\psi}_{\sigma}({\bf r})\hat{\psi}_{\sigma'}({\bf r}')\rangle$ has been addressed to deal with superconductivity.
It is even possible to invent a DFT with a distribution of nuclear sites $\Gamma({\bf R}_{1}, {\bf R}_{2}, \dots)$ (treated as points in the Born-Oppenheimer approximation) as an additional variable, which is named multicomponent DFT. Interestingly, by combined use of $\chi_{\sigma\sigma'}({\bf r},{\bf r}')$ and $\Gamma({\bf R}_{1}, {\bf R}_{2}, \dots)$ and extension to nonzero temperature, the phonon-mediated superconductivity is consistently described within the DFT framework, which is called DFT for superconductors.

The concept of introducing a non-interacting reference system that reproduces the charge density of the interacting system can also be extended to the time-dependent problem. 
Suppose the system is described by the time-dependent Schrodinger equation
\begin{eqnarray}
i\hbar\frac{\partial}{\partial t}|\Psi_{\rm e}(t)\rangle
=\hat{\mathcal{H}}_{\rm e}(t)|\Psi_{\rm e}(t)\rangle
\end{eqnarray}
with $\hat{\mathcal{H}}_{\rm e}(t)$ including a time-dependent external field $\varv({\bf r}, t)$. The Runge-Gross theorem states that, with a given initial state $|\Psi_{\rm e}(t_{0})\rangle$, a one-to-one correspondence between the time-dependent charge density $\rho({\bf r}, t)$ and external potential $\varv({\bf r}, t)$ holds \citep{Runge_Gross}. 
With this theorem, the time-dependent density functional theory has been established. 
Here, one can recast the original system to a non-interacting system
\begin{eqnarray}
i\hbar\frac{\partial}{\partial t}\phi_{i}({\bf r}, t)
=
\left[-\frac{\hbar^2}{2m}\frac{\partial^2}{\partial {\bf r}^2}+\varv_{\rm eff}({\bf r}, t)\right]\phi_{i}({\bf r}, t).
\end{eqnarray}
This time-dependent Kohn-Sham equation has been used to simulate ionization and excitation dynamics of atoms, molecules, and solids.

For further reading of the present topics, see, e.g., \citet{Parr_Yang}, \citet{Engel_Dreizler}, \citet{SCDFTI}, and \cite{SCDFTII}.

\section{Approximate exchange-correlation functionals}
\label{sec:XC}
Several exact formulas of the exchange-correlation energy $E_{\rm xc}$ are actually known. 
But such formulas are not useful because they involve the exact solution of the many-body Schr\"odinger equation. 
Practically, we design calculable approximate forms as explicit functionals of $\rho$ so that they comply with correct asymptotes, and apply them to general systems. 
Here we look over standard approximations in use.

%\subsection{Jacob's ladder}
\subsection{LDA, GGA, and orbital dependent functionals}
The concept of gradient expansion gives us a useful strategy for developing the approximate functional forms. In extended systems like periodic solids, nearly uniform itinerant electrons are expected to dominate the electronic properties, and its spatial variation would be treated as a perturbation. Along this line, the accuracy of the approximate forms may be improved systematically.

The exchange-correlation energy density $\varepsilon_{\rm xc}$ is introduced by the following decomposition
\begin{eqnarray}
E_{\rm xc}=\int \!  d{\bf r} \ \! \rho({\bf r})\varepsilon_{\rm xc}({\bf r}).
\end{eqnarray}
Here, $\varepsilon_{\rm xc}$ is in principle a functional of the entire distribution of $\rho({\bf r})$. 
A systematic approximation is implemented by expressing the spatial dependence of $\rho$ in terms of the local values (semilocal approximation)
\begin{eqnarray}
\varepsilon_{\rm xc}[\rho]({\bf r})
=\varepsilon_{\rm xc}(\rho({\bf r}), \nabla\rho({\bf r}), \dots). 
\end{eqnarray}

The first approximate form is the local density approximation (LDA)
\begin{eqnarray}
\varepsilon_{\rm xc}[\rho]({\bf r})
\simeq\varepsilon_{\rm xc}(\rho({\bf r})). 
\label{eq:LDA}
\end{eqnarray}
This form becomes exact in the uniform electron gas, where $\rho({\bf r})$ is constant in space. 
Useful approximate models within the LDA can be derived from accurate calculations for uniform electron gas by sophisticated wavefunction methods such as the diffusion Monte Carlo method.
The parameters in the models are determined by fitting the reference numerical data of $\varepsilon_{\rm xc}(\rho_{\rm uni})$ as functions of the uniform density $\rho_{\rm uni}$.
Such models, being accurate for the uniform systems, are applied to non-uniform systems with Eq.~(\ref{eq:LDA}).

The approximation can be improved by taking into account the higher order derivatives. The generalized gradient approximation (GGA) incorporates the first-order derivative of $\rho$
\begin{eqnarray}
\varepsilon_{\rm xc}[\rho]({\bf r})
\simeq\varepsilon_{\rm xc}(\rho({\bf r}),\nabla\rho({\bf r})).
\label{eq:GGA}
\end{eqnarray}
The meta-GGA incorporates the Laplacian of $\rho$, as well as kinetic energy density $\tau({\bf r})=\sum_{i}|\nabla\phi_{i}({\bf r})|^2$
\begin{eqnarray}
\varepsilon_{\rm xc}[\rho]({\bf r})
\simeq\varepsilon_{\rm xc}(\rho({\bf r}),\nabla\rho({\bf r}),\Delta\rho({\bf r}),\tau({\bf r})). 
\label{eq:metaGGA}
\end{eqnarray}
A common strategy for implementing the practical forms for those approximations is as follows: 
(i) derive asymptotic behavior of the exact functional in extreme cases, (ii) design an analytic model so that those asymptotes are reproduced, and (iii) determine the remaining model parameters, referring to some ``norm" systems or any principle like maximal smoothness.

Adding the higher-order gradients may not be efficient for incorporating quantum effects such as the Pauli exclusion and dynamical/static correlations. 
To describe such effects, the Kohn-Sham orbitals, which are also implicit functionals of $\rho$, are explicitly included as variables. 

For the former, the exact exchange term (EXX)
\begin{eqnarray}
\hspace{-0.6cm}
E^{\rm EXX} \! 
=-\frac{e^2}{2}   \sum_{ij}^{\rm occ.} \! 
\int  \! \!  \! \int   \! d{\bf r}d{\bf r}' \phi^{\ast}_{i}({\bf r})\phi^{\ast}_{j}({\bf r}')
\frac{1}{|{\bf r} \! - \! {\bf r}'|}\phi_{i}({\bf r}')\phi_{j}({\bf r}) \nonumber 
\end{eqnarray}
is added to $E_{\rm xc}$ instead of the exchange part of the functional. The magnitude of the $E^{\rm EXX}$ term is tuned by a prefactor and/or cutoff for the Coulomb potential, considering its partial cancellation with the correlation effects. 
Functionals in this form are called hybrid functionals. 

For including the dynamical and static correlation effects, a formally exact adiabatic continuation formula~\citep{Langreth-Perdew1975} provides a useful basis
\begin{eqnarray}
\! \! \!   E_{\rm c} \! \! \!  
&=& \! \! \! 
-
\frac{e^2}{2}
\int_{0}^{1} d\lambda
\int\frac{d\omega}{2\pi}
\nonumber \\
&&  \times \! 
\int \! \!  \! \int   \! d{\bf r}d{\bf r}' \ 
\frac{\chi_{\lambda}({\bf r}',{\bf r}, i\omega)-\chi_{0}({\bf r}',{\bf r}, i\omega)}{|{\bf r}-{\bf r}'|}.
\end{eqnarray}
Here, $\chi_{\lambda}$ denotes the density-density response function of the ground state with the electron-electron Coulomb interaction scaled by a factor $\lambda$. $\chi_{0}$ is the Kohn-Sham response function. 
The $\lambda$ integral is taken with the density $\rho$ fixed.
Applying approximations to $\chi_{\lambda}({\bf r}',{\bf r}, i\omega)$, we can obtain approximate $E_{\rm c}$ formulas. 
For example, the random phase approximation is widely used to approximate $\chi_{\lambda}$, and the corresponding $E_c$ successfully describes the van der Waals effect, a representative correlation effect involving unoccupied orbitals.

For a more thorough review on the LDA, GGA, and orbital-dependent functionals, see, e.g., \citet{Engel_Dreizler} and \citet{Kummel-Kronik}.

\subsection{Notes on constructing approximate exchange-correlation functionals}

As explained above, analytically derived or exactly calculated asymptotic behavior of $E_{\rm xc}[\rho]$ in the $\rho$ space is utilized to specify the form of the approximate functionals. Still, there remains an ambiguity for general non-uniform $\rho$, where the exact references are unavailable. Full implementation of the approximate forms finally requires heuristic modeling with tunable parameters. 

One approach is to make the model for $E_{\rm xc}[\rho]$ as simple as possible. For example, in constructing the GGA-PBE (Perdew-Burke-Ernzerhof) functional~\citep{GGAPBE}, a simple smooth form has been conceived, where the parameters have been unambiguously determined so that it converges to the asymptotic formulas in the limits. A more empirical approach is to refer to data of specific systems for determining the model parameters. Experimentally observed physical quantities or those calculated with accurate wave function methods are often used for this. For example, the prefactor for $E^{\rm EXX}$ is tuned so that accurately calculated cohesive energies of several molecules are optimally reproduced. See, e.g., \citet{Head-Gordon} for a thorough review on a variety of functionals.

An attempt has been recently made to remove the ambiguity stemming from the human construction of the model form. Namely, one adopts an extremely flexible model with a huge number of parameters. The model is by design capable of representing {\it any} functionals in the infinite parameter limit, so that it can mimic the ideal energy functional $E[\rho]$ as a mapping from $\rho$ to a scalar $E$. Parameters are tuned with a large amount of reference data. Modern machine learning models and parameter tuning methods are used to implement this scheme [see, e.g., \citet{Marques2019} for a review].

\section{Electronic structure}
\label{sec:electronic_structure}
DFT unambiguously defines the quantities related to the charge and spin densities and total energy; magnetic moment, lattice constants, bulk modulus, etc. 
Relating DFT to electronic single-particle properties is a more complicated issue. 
This is mainly due to the fact that the occupation number and Kohn-Sham orbital are auxiliary concepts (see Sec.~\ref{sec_HK_KS}) and not directly related to real single particle excitations.
%due to the fundamental irrelevance of the concepts of occupation number and Kohn-Sham orbital from real single-particle excitation. 
Nevertheless, the DFT calculations are widely in practical use for quantitative comparison with experimentally observed excitation spectra. 
The DFT results (Kohn-Sham wave functions, Kohn-Sham energies, etc.) can also be utilized as a basis for other methods such as the Green's function theory. 
In this section, we review those aspects.

\subsection{Single-particle spectrum}

Single-particle excitation can be observed with spectroscopy experiments.
Among the single-particle quantities, the fundamental gap, which is the minimum energy required for manipulating one electron within the system at zero temperature, is a key quantity that distinguishes insulators from metals in solids.
%Here, we discuss the relation between the true fundamental gap and the Kohn-Sham gap.
Here we discuss the discrepancy between the apparent gap in the Kohn-Sham spectra and experimentally observed optical gap (fundamental gap), to recall attention to the interpretation of the band structures obtained within the Kohn-Sham scheme.

Within DFT, the level difference between ground states with different numbers of electrons is well defined. 
%The energy required to subtract (add) an electron to an $N$ electron system, which is the ionization energy $I$ (electron affinity $A$), is thus given by
%\begin{eqnarray}
%I(N)\!\!\!&=&\!\!\!E(N-1)-E(N),
%\\ 
%A(N)\!\!\!&=&\!\!\!E(N)-E(N+1).
%\end{eqnarray}
Then, the fundamental gap for an $N$ electron system
%which is the minimum energy required for manipulating one electron within the system,
is given by
\begin{eqnarray}
\hspace{-0.5 cm }E_{\rm gap}(N)  \! \! \! 
&=&  \! \! \!  \left (  E(N \! -\! 1)-E(N)  \right  ) -  \left  ( E(N)-E(N\!+\!1)  \right )  \nonumber \\ 
&=&  \! \! \! I(N) - A(N).
\label{eq:fundamental_gap}
\end{eqnarray}
Here, $I$ and $A$ are the ionization energy and the electron affinity, respectively. 
In this section, we indicate the dependence on the total number of electrons explicitly for clarity, 
e.g., $E(N)$ is the total energy of the $N$ electron system.  
The expression of Eq.~(\ref{eq:fundamental_gap}) is, however, inefficient because we cannot always perform calculations with $N\pm 1$ electrons: For example, in bulk systems, the total $N$ is taken to infinity.

To accommodate DFT to such cases, the extension to systems with fractional electron number $N \! +\! \delta$ $(0 \! <\! \delta \! < \! 1)$ has been beneficial. There, the chemical potential $\mu$ is introduced, and the occupation number $\{n_{i}\}$ is extended to allow fractional values. 
The state can be a mixed state with fractional weights, formed by the pure states of integer electron numbers. 
The infinitesimal variation of electron numbers can be defined with this formalism. 
From this Janak's theorem~\citep{Janak} is derived, which relates the change in total energy and the occupation numbers as
\begin{eqnarray}
\delta E =  \sum_{i}\delta n_{i}\varepsilon_{i}.
\label{eq:Janak}
\end{eqnarray}
This theorem is sometimes referred to as it gives the Kohn-Sham orbitals physical meaning.
However, one has to recall that $n_{i}$ and $\varepsilon_{i}$ are artificial variables.
Thus, the operation ``changing the number of electrons in state $i$" is not directly related to adding/subtracting actual electrons. 
Yet the frontier orbital has physical significance. 
Within the extended DFT, in the $N+\delta$ electron ground state, 
%only the occupation number of the frontier orbital $n_{\rm f}$ can have a value other than 0 or 1 ($n_{i}=1   \ \forall \varepsilon_{i}\!<\!\varepsilon_{\rm f};\ n_{i}=0 \  \forall \varepsilon_{i}\!>\!\varepsilon_{\rm f}$). 
only one orbital $\phi_{\rm f}$, which is called the frontier orbital, can have a fractional occupation number ($n_{i}=1   \ \forall \varepsilon_{i}\!<\!\varepsilon_{\rm f};\ n_{i}=0 \  \forall \varepsilon_{i}\!>\!\varepsilon_{\rm f}$). 
Integrating Eq.~(\ref{eq:Janak}) for $i={\rm f}$ from the $N$ to $N\pm 1$ electron states, we obtain:
\begin{eqnarray}
I \! \! \! &=& \! \! \! \int_{N}^{N-1} dN \varepsilon_{\rm f}(N), \\
-A \! \!  \! &=& \! \! \! \int_{N}^{N+1} dN \varepsilon_{\rm f}(N).
\label{eq:I-A-exact}
\end{eqnarray}

Another important consequence of the extended DFT is that the exact $N$ dependence of the total energy is linear between integer $N$'s. 
The fact $I=E(N-1)-E(N)\neq E(N)-E(N+1)=A$ then requires that the derivative of the total energy is discontinuous at the integer points. Therefore, the curve of $E(N)$ appears as a polygonal line with nodes at the integer points~\citep{Perdew-straight}.
With this fact, the derivative gap
\begin{eqnarray}
E_{\rm gap}^{\rm deriv}
=
\left.\frac{\partial E(N)}{\partial N}\right|_{N+\delta}
-
\left.\frac{\partial E(N)}{\partial N}\right|_{N-\delta}
\end{eqnarray}
is equated to the fundamental gap. 
Furthermore, $E_{\rm gap}^{\rm deriv}$ satisfies the following equation
\begin{eqnarray}
E_{\rm gap}^{\rm deriv}
=E_{\rm gap}^{\rm KS}+\Delta_{\rm xc}
,
\label{eq:deriv-KS}
\end{eqnarray}
where $E_{\rm gap}^{\rm KS}$ is the gap between the lowest unoccupied and highest occupied Kohn-Sham states (Kohn-Sham gap) and $\Delta_{\rm xc}$ is a spatially constant term so called derivative discontinuity $\left.\frac{\partial E_{\rm xc}(N)}{\partial N}\right|_{N+\delta}
-
\left.\frac{\partial E_{\rm xc}(N)}{\partial N}\right|_{N-\delta}$~\citep{Perdew-discont}. 
This equality enables us to relate the observed gap to the quantities defined in the $N$-electron system.

%Finally, 
When we use approximate exchange-correlation functionals, $E_{\rm gap}^{\rm deriv}$ in Eq.~(\ref{eq:deriv-KS}) calculated with DFT deviates from the true fundamental gap $E_{\rm gap}$.
Then, the fundamental gap is expressed as
\begin{eqnarray}
E_{\rm gap}
%&=&E_{\rm gap}^{\rm deriv}+\left[E_{\rm gap}-E_{\rm gap}^{\rm deriv}\right]
%\nonumber \\
%&\equiv&
 = E_{\rm gap}^{\rm KS}+\Delta_{\rm xc} + \Delta^{\rm straight}.
\label{eq:gap-approx}
\end{eqnarray}
The final term $\Delta^{\rm straight}$ represents a correction term, corresponding to the deviation of the behavior of the approximate $E(N)$ from the ideal segmented straight lines.

The formula Eq.~(\ref{eq:gap-approx}) implies some interesting facts of the Kohn-Sham gap $E_{\rm gap}^{\rm KS}$. First, even with the exact functional, it does not correspond to the fundamental gap because of the nonzero $\Delta_{\rm xc}$. Second, using the standard functionals such as those within the LDA and GGA, it departs from the fundamental gap by incorrect zero $\Delta_{\rm xc}$ and nonzero $\Delta^{\rm straight}$. 
Recently, functionals design that also takes into account Eq.~(\ref{eq:gap-approx}) has been broadly conducted. For example, minimization of $\Delta^{\rm straight}$ has been referred to as a criterion for tuning some modern functionals, which improves accuracy for electronic properties. The generalized Kohn-Sham scheme, which includes the orbital-dependent functionals like EXX, helps us to incorporate a major fraction of $\Delta_{\rm xc}$ into $E_{\rm gap}^{\rm KS}$, so that the Kohn-Sham gap better agrees with the experimental fundamental gap.

See, e.g., \citet{Parr_Yang}, \citet{MoriSanchez2008}, and \citet{Burke} for further reading of the topic in this section.

\subsection{Response function}
The general fluctuation-dissipation relationship says that the electronic response to external perturbations is equated to a property in the ground state. In fact, the response function is formulated exactly within the framework of DFT \citep{Hybertsen-Louie}.
Suppose the system is perturbed by a change of the external potential $\delta \varv({\bf r})$ and the charge density distribution changes by $\delta\rho({\bf r})$. The density-density response function $\chi({\bf r}, {\bf r}')$ is defined as the linear coefficient relating them:
\begin{eqnarray}
\delta\rho({\bf r})
=
\int d{\bf r}' \chi({\bf r}, {\bf r}')\delta \varv({\bf r}')
\end{eqnarray}
The response function $\chi$ is given by
\begin{eqnarray}
\chi
=
\left[
1-\chi_{0}\left(\frac{e^2}{|{\bf r}-{\bf r}'|}
+K_{\rm xc}\right)
\right]^{-1}\chi_{0}
\end{eqnarray}
with $K_{\rm xc}({\bf r}, {\bf r}')\equiv \delta^2 E_{\rm xc}/[\delta\rho({\bf r})\delta\rho({\bf r}')]$. Here, the quantities are matrices with indices ${\bf r}$ and ${\bf r}'$. The Kohn-Sham response function $\chi_{0}$ is defined by
\begin{eqnarray}
\delta \rho({\bf r})=
\int d{\bf r}'\chi_{0}({\bf r},{\bf r}')\delta \varv_{\rm eff}({\bf r}'). 
\end{eqnarray}
Applying the perturbation theory to the Kohn-Sham equation, 
we can write $\chi_{0}$ in terms of the Kohn-Sham eigenpairs ($\varepsilon_{i}$, $\phi_{i}$). Note that those expressions are rigorous and well-defined within DFT, though all the ambiguities are condensed in $E_{\rm xc}$. Using the time-dependent DFT (Sec.~\ref{sec:extension}), the time-dependent response function can also be formulated.

\subsection{Kohn-Sham states in practice}
DFT is in principle independent from the Green's function theory since the former does not define the electron one-particle creation/annihilation operator. Nevertheless, practices are found in the literature where the Kohn-Sham eigenstates are utilized as a basis for calculating quantities formulated in the Green's function theory, showing remarkable successes in comparing with experiments. Calculations of this kind may be justified by the fact that the Green's function theory can be made basis-free when the perturbation effects are incorporated up to the infinite order \citep{Hedin1965}. The Kohn-Sham states, which are introduced as auxiliary quantities (see Sec.~\ref{sec_HK_KS}), are then presumed as a good zeroth-order basis, which enables us relatively fast convergence with respect to the order of perturbation.

\begin{figure}[tbp]
\vspace{0cm}
\begin{center}
\includegraphics[width=0.45\textwidth]{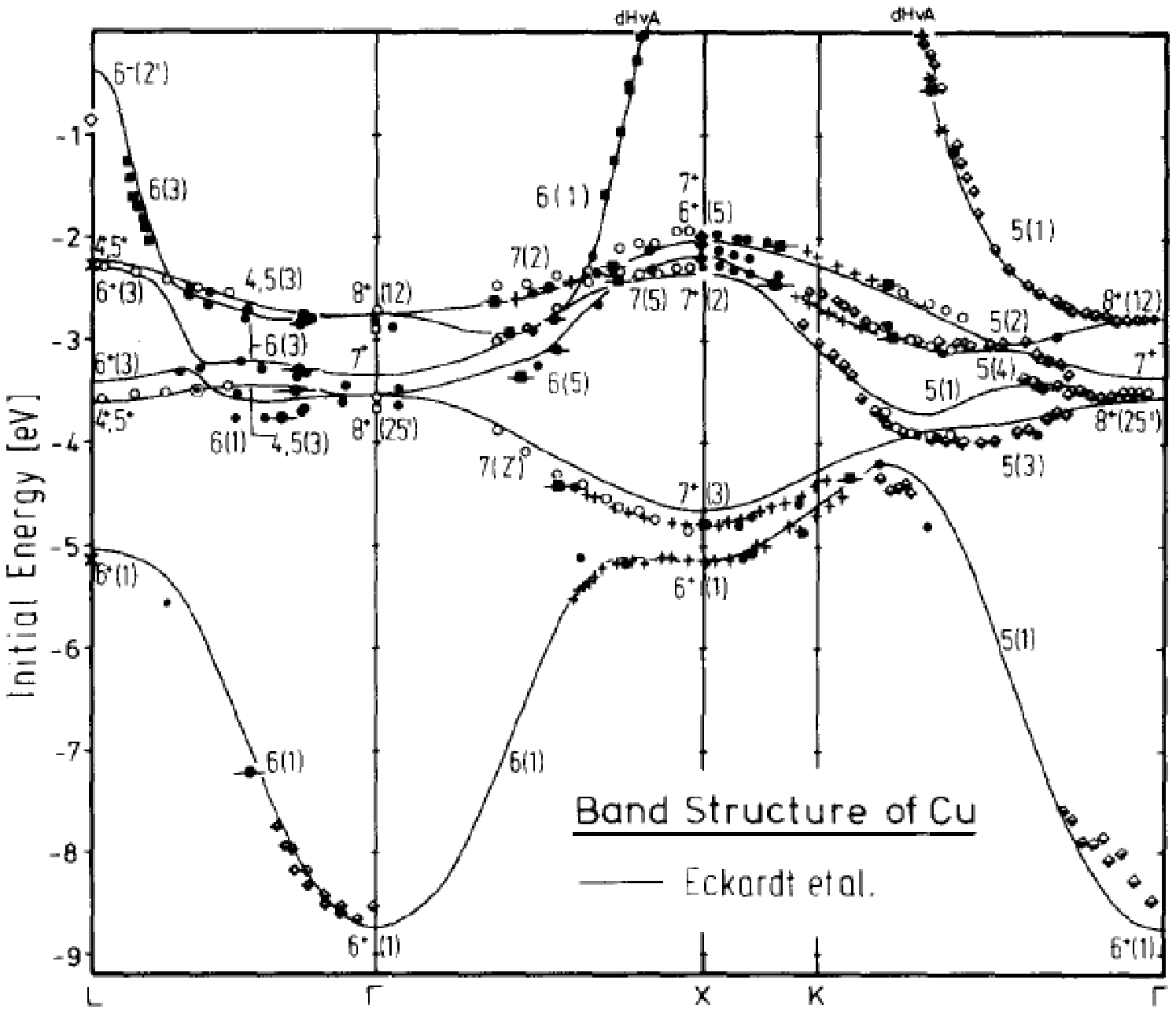}
\caption{Electronic band structure of copper calculated by the Kohn-Sham equation with a primary LDA functional. The points are from photoemission experiments. 
%Reprinted from Physics Reports \textbf{112}, R. Courths and S. H\"ufner, "Photoemission experiments on copper", 53, Copyright (1984), with permission from Elsevier.
Reproduced from \citet{Courths_Hufner}, with permission from Elsevier.
}
\label{Fig_elband}
\end{center}
\end{figure}

\begin{figure}[tbp]
\vspace{0cm}
\begin{center}
\includegraphics[width=0.45\textwidth]{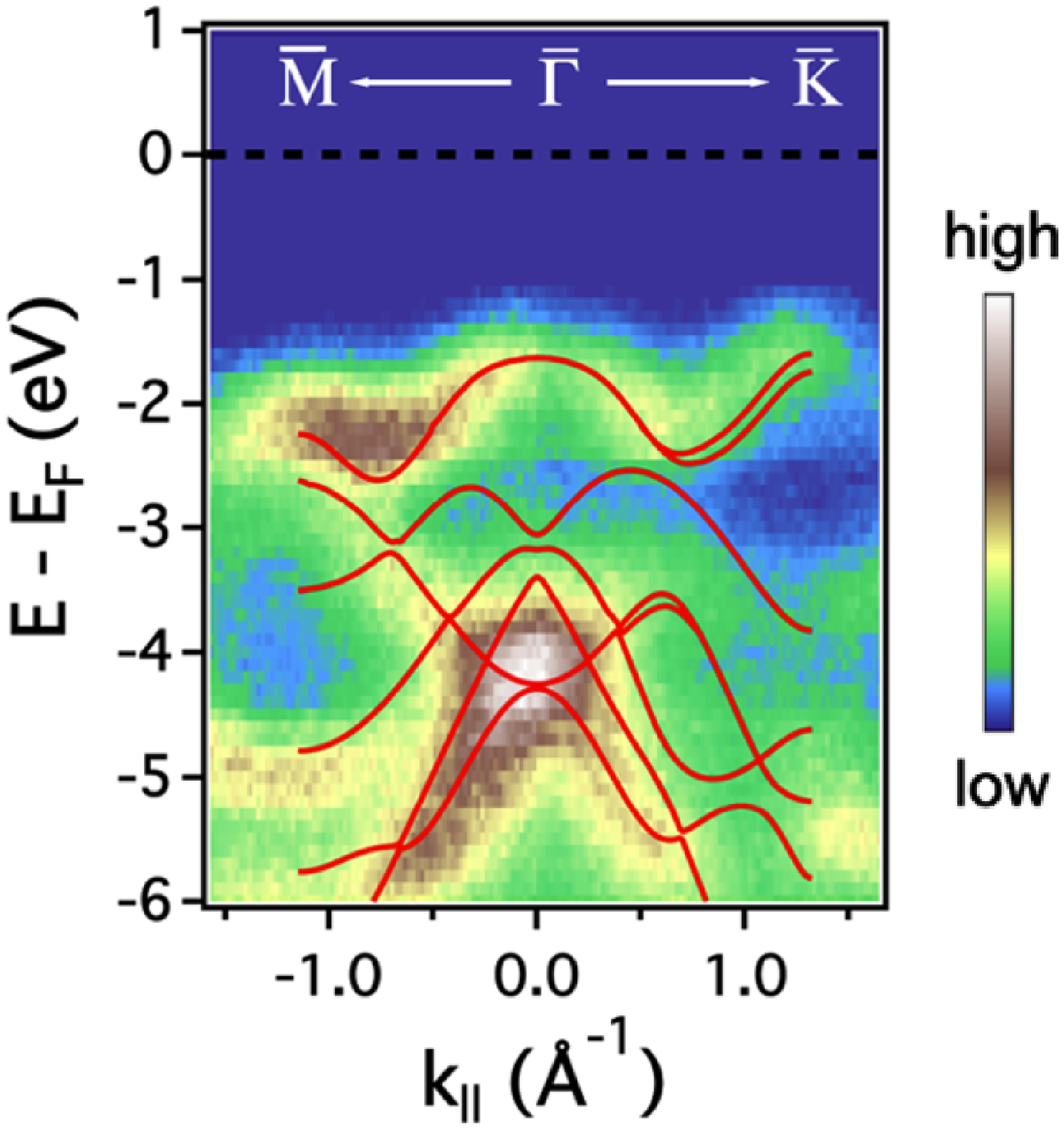}
\caption{
Intensity map of the photoemission spectrum of exfoliated monolayer MoS$_{2}$ measured with micrometer-scale ARPES.  
The GGA-PBE Kohn-Sham band structure including spin-orbit interaction \citep{Z_Zhu_2011} is indicated by the red curves.
Reproduced with permission from \citet{Jin_2013}.
Copyright (2013) by the American Physical Society.
}
\label{Fig_elband2}
\end{center}
\end{figure}

The combination of DFT and the Green's function theory has yielded tremendous success in describing the single-particle excitation of materials. 
The spectral function defined in the Green's function theory is directly measured by, e.g., angle-resolved photoemission spectroscopy (ARPES). 
Although the agreement between the Kohn-Sham and experimental spectra is not theoretically guaranteed, a number of studies have conducted comparisons between them. 
A classic application to copper is displayed in Fig.~\ref{Fig_elband}. 
Recently, the comparison serves more and more essential roles in analyzing the detailed electronic structure of materials, with the improvement in the experimental precision: Here, we also append an example of monolayer MoS$_2$ in Fig.~\ref{Fig_elband2}. 
These theoretical spectra would need to be interpreted as results of the Green's function theory using the Kohn-Sham states as the zeroth-order basis.

The Kohn-Sham calculation is known to yield reasonable spectra in weakly correlated systems such as dense metals.
However, the agreement with experiments becomes worse for insulators (especially their band gap) and dilute metals, where the exchange-correlation effects are stronger. Even in such cases, accuracy can be efficiently cured by introducing the partially screened exchange term to the functional or calculating the self-energy by perturbation theory with the Kohn-Sham eigenbasis. The GW approximation [see, e.g., \citet{Aryasetiawan_Gunnarsson} for a review] serves as a standard method for calculating the self-energy: It treats the long-range exchange-correlation effects efficiently under weak to strong Coulomb interaction.

When the correlation effects become even stronger, the one-electron description may break down. 
Such a situation occurs in so-called strongly-correlated materials, where energy scales of the interaction and kinetic energies compete. 
Typical strongly-correlated materials are, for example, transition-metal oxides with partially-filled $d$-electron shells and heavy-fermion materials with partially-filled $f$-electron shells. 
The bandwidth $W$ of partially-filled orbitals in these materials is typically narrow, and becomes comparable or even smaller than the onsite Coulomb interaction $U$.
For such systems, the band theory based on the KS states becomes inaccurate, and it often fails to describe the spectral property around the Fermi level. 
For example, when we regard the Kohn-Sham energy eigenvalues as the poles of the spectral function, DFT cannot reproduce the charge gap opening in Mott insulators, where electrons are localized due to the strong Coulomb repulsion.
To incorporate such many-body effects, a combination of DFT and many-body methods, such as the dynamical mean-field theory and the variational Monte Carlo method, has been developed. 
For more details of the conceptual and practical aspects of the combination, see, e.g., \citet{Kotliar_2006} and \citet{Imada_2010}.

\section{Nuclear dynamics}
\label{sec:lattice}

DFT is also useful in investigating nuclear dynamics. 
This is because the total energy of the electronic system plays a role as the potential for the atomic vibration [Eq.~(\ref{Eq:lattice})]. 
Here, we discuss the expressions for the forces acting on nuclei and the interatomic force constants, which are used in structure optimization and phonon dispersion calculations, respectively.

\subsection{Forces acting on nuclei}

The force acting on $I$th nucleus ${\bf F}_I$ is given by the derivative of the energy surface
\begin{eqnarray}
{\bf F}_{I} =  -\frac{\partial E \bigl( \{ {\bf R} \} \bigr )}{\partial {\bf R}_{I}}. 
\end{eqnarray}
The forces on nuclei vanish (${\bf F}_{I} = {\bf 0}$) at equilibrium geometry.
Therefore, one can perform structure optimization by minimizing the forces on the nuclei.  
In practice, the forces (first derivative of the energy surface) can be estimated efficiently using the Hellmann-Feynman theorem [\citet{Hellmann}, \citet{Feynman}].

\subsection{Interatomic force constants}

Here, we show how the normal vibrational modes are derived. 
We consider the displacement of $I$th nucleus ${\bf u}_I$ from the equilibrium position ${\bf R}_{I}^{(0)}$. 
The position of the $I$th nucleus reads
\begin{eqnarray}
{\bf R}_{I}={\bf R}_{I}^{(0)}+{\bf u}_{I},
\end{eqnarray}
Then, the kinetic energy $T$ for the atomic vibration is given by 
\begin{eqnarray}\label{Eq:T_kin}
T = \frac{1}{2}\sum_{I }M_{I}|\dot{{\bf u}}_{I}|^{2} = 
\frac{1}{2}\sum_{I  \alpha}M_{I}(\dot{ u}_{I}^{\alpha})^{2}.
\end{eqnarray}
Here, $\alpha$ is the cartesian component of the displacement ($\alpha = x,y,z$).
For the the potential energy $U$, we apply the harmonic approximation:
\begin{eqnarray}
\label{Eq:U_pot}
U &=& E(\verb|{| {\bf R}^{(0)} + {\bf u} \verb|}|)-E(\verb|{| {\bf R}^{(0)} \verb|}| ) \\
&\simeq& \frac{1}{2} \sum_{I \alpha} \sum_{I^{\prime} \alpha^{\prime}}  
\frac{\partial^{2} E \bigl( \{ {\bf R} \} \bigr ) }{\partial  R_{I}^{\alpha} \partial R_{I^{\prime}}^{\alpha^{\prime}}} 
u_{I}^{\alpha}u_{I^{\prime}}^{\alpha^{\prime}}. 
\end{eqnarray}
Note that the first-order terms with respect to the displacement vanish at the equilibrium geometry.

From the expression of $U$ and $T$, one can derive the secular equation, which determines the frequency and displacement pattern of the normal vibrational modes: 
\begin{eqnarray}
\label{Eq:motion_R}
\sum_{I^{\prime}  \alpha^{\prime}} \left(  C_{I, I^{\prime}}^{\alpha \alpha^{\prime}}  - M_I \omega^2  \delta_{I I^{\prime}}  \delta_{\alpha \alpha^{\prime}} \right) u_{I^{\prime}}^{\alpha^{\prime}}  = 0 ,
\end{eqnarray}
where $C$ is the matrix of so called interatomic force constants \begin{eqnarray}
C_{I, I^{\prime} }^{\alpha \alpha^{\prime}}=
\frac{\partial^{2} E \bigl( \{ {\bf R} \} \bigr ) }{\partial R_{I}^{\alpha} 
\partial R_{I^{\prime}}^{\alpha^{\prime}}}.
\end{eqnarray}
The interatomic force constants play a role as ``spring constants'' of the ``springs'' between nuclei.
In crystals, Eq.~(\ref{Eq:motion_R}) becomes block diagonal in momentum space, and the normal modes are labelled by wave vectors. 
Then, the frequencies of the normal modes give the phonon dispersion in solids. 

In practical calculations, there exist mainly two approaches to estimate the interatomic force constants. 
One is a direct method, called frozen phonon, in which forces are computed under finite amplitudes of displacements and the interatomic force constants are estimated from finite differences of the forces. 
The other method relies on the density-functional perturbation theory (DFPT)~\citep{Baroni_2001}, which shows that the interatomic force constants can be computed from the ground-state electron density and its linear response to the atomic displacements. 
Therefore, the frozen phonon method performs supercell calculations with finite atomic displacements, whereas the DFPT performs linear response calculations to the displacements.

\begin{figure}[tbp]
\vspace{0cm}
\begin{center}
\includegraphics[width=0.48\textwidth]{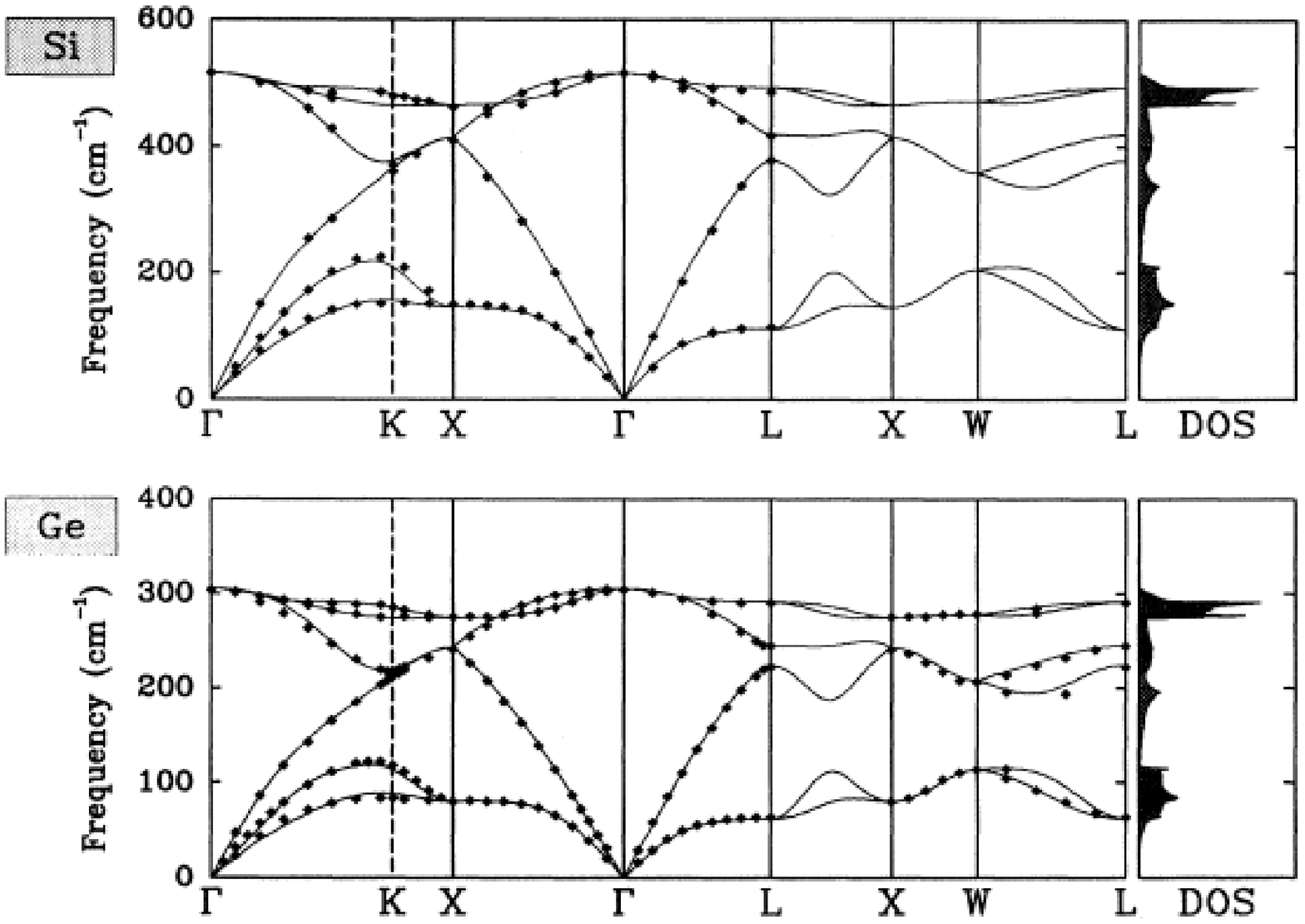}
\caption{
Phonon dispersions and densities of states of silicon and germanium solids. The diamonds are experimental data. 
Reproduced with permission from \citet{Giannozzi_1991}. Copyright (1991) by the American Physical Society.
}
\label{Fig_phonon}
\end{center}
\end{figure}

As an example of phonon calculations, we show, in Fig.~\ref{Fig_phonon}, the phonon dispersions of elemental semiconductors, Si and Ge, calculated using the DFPT~\citep{Giannozzi_1991}.
The DFPT results show a good agreement with experiments.

\subsection{Recent topics}

Here, among many recent topics, we introduce two advances related to phonon properties: anharmonicity and structure prediction.  

Anharmonicity is a deviation of the atomic vibrations from a harmonic oscillator, which arises from higher-order terms of the expansion of energy difference with respect to atomic displacements [Eq.~(\ref{Eq:U_pot})].
The anharmonic terms are crucial in describing thermal properties of atomic vibrations, such as thermal expansion, thermal conductivity, and thermodynamic stability. 
Optimizing these properties leads to functional materials. 
For example, controlling the phonon lifetime and achieving low thermal conductivity in the phonon-contributing part is essential in designing good thermoelectric materials. 
For more details of the recent advancements in the computation of anharmonicity, see, e.g., \citet{Tadano_2018,McGaughey_2019}.

One of the dreams of materials scientists would be a theoretical prediction of functional materials. 
Structure prediction of materials is a very challenging task because the energy landscape $E \bigl( \{ {\bf R} \} \bigr )$ is a complicated object with a huge number of local minima. 
Recently, there has been a tremendous advance in numerical tools to explore the complex potential energy surface 
(see, e.g., \citet{Andreoni_2020}). 
A remarkable success of the crystal-structure search algorithm so far is, for example, a prediction of high-temperature superconductors under extremely high pressure (see, e.g., \citet{Flores-Livas_2020} for details).

\section{Practical guide}
\label{sec:practical}

\subsection{Program packages}

Nowadays, there are a variety of DFT open source packages.
Some of them are introduced in \citet{Talirz_2021}. 
See also, e.g.,
\url{https://en.wikipedia.org/wiki/List_of_quantum_chemistry_and_solid-state_physics_software}
for a list.

\subsection{Reproducibility}

In the practical calculations of DFT, most of the open-source programs employ the Kohn-Sham formalism (see Sec.~\ref{sec_HK_KS}).
When the same exchange-correlation functional is employed, different programs solve the same Kohn-Sham equations. 
Ideally, all codes should give the same solution.
However, depending on, for example, the type of the basis set [plane-wave, (linearized) augmented plane wave (L)APW, linear muffin-tin orbital (LMTO), and so on] and approximate ionic potentials [norm-conserving and ultrasoft pseudopotentials, projector augmented wave (PAW) method, and so on], the accuracy may differ among available packages. 
Therefore, it is an important task for the community to check the reproducibility of the DFT calculations. 
Recently, such a systematic benchmark has started to be performed. 
For example, \citet{Lejaeghere_2016} compared various DFT packages for solids and confirmed that widely-used codes and methods give essentially identical solutions. 

\section{Summary}
\label{sec:summary}

DFT is currently one of the most standard and established methods for electronic structure calculations.
Many DFT open sources have been developed in recent years, allowing a wide range of researchers to enjoy DFT calculations.
Nowadays, DFT is widely used to calculate the properties of solids. In addition to the ground-state electronic structure calculations, DFT is also useful for calculating excitation spectra (both single- and two-particle) and nuclear dynamics. 
Care must be taken then in interpreting the eigenvalues of the Kohn-Sham equation.

Although it is impossible to cover all of the vast DFT-related topics, we have provided an introductory explanation of some of them in this chapter.
Please refer to the references for more details of advanced topics.
We hope this chapter will be of some help to understand what can be done with DFT.

\section*{Acknowledgments}

We are grateful for valuable comments from Hideo Aoki, Ryotaro Arita, Silke Biermann, Kieron Burke, Kenta Kuroda, Yasushi Shinohara, Terumasa Tadano, and Shinji Tsuneyuki. 

%\appendix

%\section{Sample Appendix Section}


\bibliographystyle{elsarticle-harv} 
\bibliography{cas-refs}





% \end{thebibliography}
\end{document}